\documentclass[10pt]{article}
\usepackage{amsmath,euscript,amssymb}
\usepackage{pxfonts}
\usepackage{mathpazo}
\usepackage[dvips]{graphicx}
\usepackage[T1]{fontenc}
\usepackage{epsfig}
\usepackage[cp1251]{inputenc}
\usepackage{longtable,supertabular}
\def\v1{\vspace{1cm}}
\def\be{\begin{equation}}
\def\ee{\end{equation}}
\def\bc{\begin{center}}
\def\ec{\end{center}}

\def\vh{\varphi}

\newcommand{\bea}{\begin{eqnarray}}
\newcommand{\eea}{\end{eqnarray}}

\begin{document}
\title{\textbf{\textsc{Hamiltonian General Relativity in CMB frame}} \\[1cm]}
 \author{B.M. Barbashov${}^{1}$, \L.A. Glinka${}^{1,2}$\footnote{\textbf{Electronic addresses: glinka@theor.jinr.ru, lukaszglinka@wp.eu}},
 V.N. Pervushin${}^{1}$,
 and A.F. Zakharov${}^{1,3,4}$
 \\[1cm]
{\normalsize\it $^1$ Bogoliubov Laboratory of Theoretical Physics,
Joint
Institute for Nuclear Research,}\\
{\normalsize\it 6 Joliot--Curie Street, 141980, Dubna, Russia}\\
{\normalsize\it $^2$ College of Inter-faculty Individual Studies in Mathematics and Natural Sciences (MISMaP),}\\
{\normalsize\it Warsaw University, 93 Zwirki i Wigury Street, Room 156, 02-089, Warsaw, Poland}\\
{\normalsize\it $^3$ National Astronomical Observatories of Chinese
Academy
of Sciences, Beijing 100012, China}\\
{\normalsize\it $^4$ Institute of Theoretical and Experimental
Physics, 25, 117259, Moscow, Russia} }

\date{\empty}
\maketitle
\medskip

\begin{abstract}
 {\noindent

A collection of requirements to the General Relativity that follow
from the WMAP observations of the Cosmic Microwave Background
radiation anisotropy as an inertial frame are discussed. These
obligations include the  separation of both the CMB frame from the
\mbox{diffeomorphisms} and the diffeo-invariant cosmic evolution
from the local scalar metric component in the manner compatible with
the canonical Hamiltonian approach to the Einstein--Hilbert theory
with the energy constraints. The solution of these constraints in
classical and quantum theories and a fit of units of measurements
are discussed in the light of the last Supernovae data.}
\end{abstract}

\newpage

\tableofcontents

\section{Introduction}

 Measurement of the dipole component of Cosmic Microwave Background (CMB)
 radiation temperature $T_0(\theta)=T_0[1+(\beta/c)\cos\theta]$, where
 $\beta=390\pm 30$ km/s,  \cite{WMAP} testifies to motion of the Earth to
 the Leo with the velocity $|\vec v|$= 390$\pm$ 30 km/s with respect to CMB,
 where 30 km/s rejects the copernican annual motion of the Earth
 around the Sun, and 390 km/s to the Leo is treated as the
 parameter of the Lorentz transformation from the the Earth frame to the
 CMB frame. The CMB frame can be identified with the comoving
 {\it inertial frame} of
 the Early Universe created at zero moment of its {\it proper time},
 if  the CMB is considered  as the evidence of such the creation.

 This relativistic treatment of the observational data produces the definite
 questions to the General Theory of
 Relativity and the modern cosmological models destined for description
 of the processes  of origin of the Universe and its evolution:

\begin{enumerate}
\item How can separate the CMB frame from the
general coordinate transformations?
\item How can separate the cosmic evolution
from a dynamics of the local scalar component in the CMB reference
frame?
\item What are the requirements of the CMB data description
to the canonical approach to the General Relativity and the Standard
Model including the Vacuum Postulate?
\end{enumerate}

 In this paper, we try to get the possible responses
 to these issues that follow from the principles of General Theory of Relativity and Quantum Field Theory, in order to clear up
 restrictions of the cosmic motion of the Universe in
 both the Minkowski space of events and in the Wheeler-DeWitt field space
 of events \cite{WDW}.

 We show how these responses can help to clear up the origin of CMB
 and to explain the energetical budget of our Universe.

\section{Canonical General Relativity}

\subsection{The Fock separation of the frame transformations
 from diffeomorphisms}

   Recall that the Einstein--Hilbert theory of gravitation
    is given by two fundamental quantities, they are a
 {\it geometric interval}
\be \label{1-2}
 ds^2=g_{\mu\nu}dx^\mu dx^\nu
 \ee
   and  the  {\it dynamic} Hilbert action
 \be\label{1-1}
 S[\vh_0|F]=\int d^4x\sqrt{-g}\left[-\frac{\vh_0^2}{6}R(g)
 +{\cal L}_{(\rm M)}\right]=\int d^4x \sqrt{-g}{\cal L},
 \ee
 where $\varphi^2_0=\dfrac{3}{8\pi}M^2_{\rm Planck}=
 \dfrac{3}{8\pi G}$, $G$ is the Newton
 constant in the units $\hbar=c=1$,
 ${\cal L}_{(\rm M)}$  is the  Lagrangian of the matter fields. This action is clearly dependent on the collection of fields and metric
 $F=[f,g]$.
 These fundamental quantities (\ref{1-2}) and (\ref{1-1}) are invariant with respect to action
 of general coordinate transformations, known widely as diffeomorphisms
 \be \label{1-5}
 x^{\mu} \rightarrow  \tilde x^{\mu}=\tilde
 x^{\mu}(x^0,x^{1},x^{2},x^{3}),
 \ee
 Separation of  the diffeomorphisms from
 the Lorentz transformations in GR is fulfilled
 by  introduction of a square root
 of the interval \cite{fock29}
\be \label{1-3}
 ds^2\equiv\omega_{(\alpha)}\omega_{(\alpha)}=
 \omega_{(0)}\omega_{(0)}-
 \omega_{(1)}\omega_{(1)}-\omega_{(2)}\omega_{(2)}-\omega_{(3)}\omega_{(3)},
 \ee
 where $\omega_{(\alpha)}$  are linear differential forms
 invariant with respect to action of diffeomorphisms
 $$
 \omega_{(\alpha)}(x^{\mu})~\to ~\omega_{(\alpha)}(\tilde x^{\mu})=
 \omega_{(\alpha)}(x^{\mu}).
 $$
  These forms are treated
as components of an orthogonal
 simplex of reference with the following Lorentz transformations
\bea \label{1-4} {\omega}_{(\alpha)}~\to
~\overline{\omega}_{(\alpha)}=
\overline{\omega}_{(\alpha)}=L_{(\alpha)(\beta)}{\omega}_{(\beta)}.
\eea

 There is an essential difference between diffeomorphisms (\ref{1-5}) and
 the Lorentz
 transformations  (\ref{1-4}). Namely,
 parameters of the Lorentz
 transformations  (\ref{1-4}) are measurable quantities, while parameters
 of diffeomorphisms (\ref{1-5}) are unmeasurable one. Especially, the simplex
 components ${\omega}_{(\alpha)}$ in the Earth frame
   moving with respect
  to Cosmic Microwave Background (CMB) radiation with
 the velocity $|\vec v|$= 390 km/s to the Leo
 are connected with the simplex
 components in the CMB frame $\overline{\omega}$ by the following formulae
 \bea \label{1-6} \overline{\omega}_{(0)}&=&\frac{1}{\sqrt{1-\vec
 v^2}}\left[\omega_{(0)}- v_{(c)}\omega_{(c)}\right],\\\nonumber
 \overline{\omega}_{(b)}&=&\frac{1}{\sqrt{1-\vec
 v^2}}\left[\omega_{(b)}- v_{(b)}\omega_{(0)}\right], \eea
 where the velocities $\vec v$ are measured \cite{WMAP} by the the modulus of
 the dipole component
  of CMB temperature $T_0(\theta)=T_0[1+(\beta/c)\cos\theta]$
   and its direction at the
   space\footnote{The invariance of
   the action with respect to
 frame transformations means that there are
 integrals of motion (the first Noether theorem \cite{Noter});
 while the invariance of the action with respect to diffeomorphisms
 leads to the Gauss type constraints between the motion integrals
 (the second Noether theorem \cite{Noter}). These constraints are derived in a {\it specific frame
 of reference to the initial data}. The constraints mean that only
 a part of metric components becomes {\it degrees of freedom} with the initial data.
 Another part corresponds to the diffeo-invariant {\it static  potentials}
 that does not have initial data because their equations
 contain the Beltrami-Laplace operator. And the third part of
 metric components after the resolution of constraints becomes
 diffeo-invariant non-dynamical variables that can be excluded
 by the gauge-constraints \cite{dir} like the longitudinal fields in Quantum Electrodynamics \cite{d}.}.

\subsection{The Dirac -- ADM canonical General Theory of Relativity in the CMB frame}

 The problem of the choice of a specific frame destined for description of
 evolution of the
 Universe in GR  was formulated by Dirac  and Arnovitt,
 Deser and Misner \cite{dir} as 3+1 foliated space-time (see also \cite{vlad}).
 This foliation can be
 rewritten in terms of the Fock simplex components as follows
 \be
\label{1-7}
 \omega_{(0)}=\psi^6N_{\rm d}dx^0,
 ~~~~~~~~~~~
 \omega_{(b)}=\psi^2 {\bf e}_{(b)i}(dx^i+N^i dx^0),
 \ee
 where triads ${\bf e}_{(a)i}$ form the spatial metrics with $\det |{\bf
 e}|=1$, $N_{\rm d}$ is the Dirac lapse function, $N^k$ is the shift
 vector and $\psi$ is a determinant of the spatial metric.

 In this case, the accepted canonical  Dirac -- ADM approach \cite{dir} to GR
 is given by the action  (\ref{1-1})
   in the Hamiltonian form
 \be\label{h-1}
 S_{\rm canonical}[\vh_0|F]=\int dx^0\int d^3x
 \left(\sum\limits_{{F}
 } P_{F}\partial_0F
 +{\cal C}-{N_d} T_{\rm d}\right),
 \ee
 where
 $P_F=(p_{\psi},p^i_{(b)},P_f)$ is the set of
   canonical momenta including the metric component ones
\bea\label{m-1}{p_{\psi}}&=&\frac{\partial [\sqrt{-g}{\cal
L}]}{\partial
 (\partial_0\ln{{{\psi}}})}=-\frac{8\vh_0^2}{{N_d}}\left[
 (\partial_0-N^l\partial_l)\log{
 {\psi}}-\frac16\partial_lN^l\right],
 \\\label{m-2}
 p^i_{(b)}&=&\frac{\partial[\sqrt{-g}{\cal L}]
 }{\partial(\partial_0{\bf e}_{(a)i})}
 ={\bf e}^i_{(a)}\frac{\vh_0^2}{6}\left({\bf e}_{(a)i}v^i_{(b)}+{\bf
 e}_{(b)i}v^i_{(a)}\right),
\label{proizvod1} \eea
 where
 \be
 v_{(a)i}=
 \frac{1}{{N_d}}\left[(\partial_0-N^l\partial_l){\bf e}_{(a)i}
+ \frac13 {\bf
 e}_{(a)i}\partial_lN^l-{\bf e}_{(a)l}\partial_iN^l\right],
 \ee
and
 \be\label{h-2}
 {\cal C}=N_{(b)}
  {T}^0_{(b)} +\lambda_0{p_\psi}+ \lambda_{(a)}\partial_k{\bf e}^k_{(a)}
 \ee
 is a sum of the constraints with the Lagrangian multipliers, including
 three  first class constraints
\bea\label{h-c1}
 -{\bf e}^k_{(a)}\dfrac{\delta S}{\delta N^k}={T^0_{(a)}}
 =
  -p_{\psi}\partial_{(a)}
 {\psi}+\frac{1}{6}\partial_{(a)}
 (p_{\psi}{\psi}) +
 2p_{(bc)}\sigma_{(b)|(a)(c)}-\partial_{(b)}p_{(ba)}
  +{T^0_{(a)}}_{({\rm m})}~ ,
  \eea
 where ${p}_{(ab)}=\dfrac{1}{2}({\bf e}^k_{(a)}
 {p}_{(b)k}+
 {\bf e}^k_{(b)}{p}_{(a)k})$,
 $
 \sigma_{(a)|(b)(c)}=
 {\bf e}_{(c)}^{j}
 \nabla_{i}{\bf e}_{(a) k}{\bf e}_{(b)}^{\phantom{r}k}=
 \frac{1}{2}{\bf e}_{(a)j}\left[\partial_{(b)}{\bf e}^j_{(c)}
 -\partial_{(c)}{\bf e}^j_{(b)}\right]
  $
  are the coefficients of the spin-connection (see \cite{ll} Eq.
  (98.9)),
  and four the second class ones \cite{dir}
 \bea\label{h-l1}
 \partial_k{\bf e}^k_{(a)}&=&0,\\\label{h-l2}
 p_\psi&=&0.
 \eea

 It is not difficult to check that the last constraints, i.e.  the zero momentum
  of the spatial volume element
\be\label{h-c2}
 p_{{\psi}}=-8\vh_0^2{v_{\psi}}=0 \to
 \partial_0({\psi}^6)
 =\partial_l({\psi}^6 N^l),
 \ee
   means the minimal hypersurface of
  imbedding of the three-dimensional manifold into four-dimensional
  space-time,  and these constraints lead to the
   Hamiltonian density
\be\label{h-3}
 -\frac{\delta S}{\delta N_{\rm d}}\equiv T_{\rm d}=
 \dfrac{4\varphi_0^2}{3}{\psi}^{7}
 \triangle_{\mathrm{BL}}
{\psi}+
  \sum\limits_{I=0,4,6,8,12} {\psi}^I{\cal T}_I=0,
 \ee
 where
\be\label{h-4} \triangle_{\mathrm{BL}}
 {\psi}\equiv\frac{1}{\sqrt{\gamma}}\big(\partial_{(a)}
 \sqrt{\gamma}\gamma^{\rm{(ab)}}\partial_{(b)}\big)\psi=\partial_{(b)}\partial_{(b)}{\psi}
 \ee
 is
 the Beltrami--Laplace operator, $\gamma^{(ab)}$ is a spatial
 metric,
 $\partial_{(a)}={\bf e}^k_{(a)}\partial_k$, and
 ${\cal T}_I$ is partial Hamiltonian density
  marked by the index $I$ running a collection of values
   $I=0$ (stiff), 4 (radiation), 6 (mass), 8 (curvature), 12
   ($\Lambda$-term)
in accordance with a type of matter field contributions,
 particularly for metric components these densities take the
 following form \cite{bpzz}
 \bea
\label{h-5}
 {\cal T}_{I=0}&=&\dfrac{6}{\vh_0^2}{p}_{(ab)}^{2}
 -\dfrac{16}{\vh_0^2}p^2_{\psi},
 \\\label{h-6}
 {\cal T}_{I=8}&=&\dfrac{\varphi_0^2}
  {6}~\,{}^{(3)\!}R({\bf e}),
\eea
 where \be \label{1-17}
 {}^{(3)}R({\bf e})=-2\partial^{\phantom{f}}_{i}
 [{\bf e}_{(b)}^{i}\sigma_{{(c)|(b)(c)}}]-
 \sigma_{(c)|(b)(c)}\sigma_{(a)|(b)(a)}+
 \sigma_{(c)|(d)(f)}^{\phantom{(f)}}\sigma^{\phantom{(f)}}_{(f)|(d)(c)}
 \ee
is a spatial
 curvature.

\subsection{The Lichnerowicz variables and {\it relative units}
of the dilaton gravitation}

The dependence on the energy momentum tensors (\ref{h-3})
    on the
   spatial determinant {\it potential} $\psi$ is completely determined by
   the Lichnerowicz (L) transformation to the
   conformal variables
  \bea \label{1-12}
 \omega_{(\mu)}&=&\psi^{2}\omega^{(L)}_{(\mu)},\\\label{1-13}
 g_{\mu\nu}&=&\psi^{4}\,g_{\mu\nu}^{(L)} ,\\\label{1-14}
 F^{(n)}&=&\psi^{-2n}\,F_{(L)}^{(n)},
 \eea
 where $F^{(n)}$ is any field with the one of conformal weights
 $(n)$: $n_{\rm scalar}=1$, $n_{\rm spinor}=3/2$, \mbox{$n_{\rm
 vector}=0$}.

 One can say that
 the manifest dependence on the energy density $T_{\rm d}$
 on the spacial determinant $\psi$ in the expression (\ref{h-1})
 is equivalent to a choice the L-coordinates (\ref{1-12}) $\omega^{(L)}_{(\mu)}$
 and L-variables
 (\ref{1-13}),  (\ref{1-14}) as observable ones.
 The L-observables are physically equivalent with the case
 when the field
  with the mass $m=m_0\psi^2$  is contained in
 space-time
 distinguished by the unit
 spatial metric determinant and the volume element
 \be \label{1-15}
 dV^{(L)}=\omega^{(L)}_{(1)}\wedge \omega^{(L)}_{(2)}\wedge
 \omega^{(L)}_{(3)} =d^3x.
 \ee
 In terms of the L-variables and L-coordinates $\vh_0\psi^2=w$
 the Hilbert action of classical theory of gravitation
  (\ref{1-1}) is formally the same as the action of
  the dilaton gravitation (DG) \cite{pct}
\be\label{dg-1}
  S_{DG}[\hat g^w]=-\int d^4x\frac{\sqrt{-\hat g^w}}{6}~R(\hat g^w)\equiv
  -\int d^4x\left[\frac{\sqrt{- g}w^2}{6}~R( g)-w
  \partial_\mu(\sqrt{- g}\partial_\nu w g^{\mu\nu})\right],
 \ee
 where $\hat g^w=w^2g$ and
   $w$ is the dilaton scalar field. This action is invariant with
 respect to the scale transformations
 \be\label{dg2}
 {F^{(n)\Omega}}=\Omega^{n}F^{(n)}, ~~{g^{\Omega}}=\Omega^{2}g,
 ~~{w^{\Omega}}=\Omega^{-1}w.
 \ee
 One can see that there is a transformation \be\label{ct1}
 \Omega=\dfrac{w}{\vh_0}\ee
 converting the dilaton action  (\ref{dg-1}) into the Hilbert one
   (\ref{1-1}).
In this manner, the CMB frame reveals the possibility to
 choose the units of measurements in the canonical  GR.

\subsection{The Newton law}

 The $\psi$-independence of L-variables are compatible with the
 Newton law and the
 cosmological dependence of the energy density on the scale factor $a$ in the homogeneous
 approximation $\psi^2\to a$.

 The Newton law is determined by the energy constraints   (\ref{h-3})
 and the equation of motion of the
   spatial determinant that,
   in the case of the minimal surface constraints (\ref{h-c2}),
    take the potential form
   \be\label{h-c3}
  -\psi\frac{\delta S}{\delta \psi}\equiv T_{\psi}=
  \dfrac{4\varphi_0^2}{3}\left\{7N_d{\psi}^{7}\triangle_{\mathrm{BL}} {\psi}+{\psi} \triangle_{\mathrm{BL}}
\left[N_d{\psi}^{7}\right]\right\}+
  N_d\sum\limits_{I=0,8}I {\psi}^I{\cal T}_I=0.
 \ee
It is not embarrassing to check that in a region of the empty space,
where
 two dynamic variables are absent ${\bf e}_{(a)k}=\delta_{(a)k}$
 (i.e. ${\cal T}_{I}=0$), one can get the Schwarzschild-type
 solution of equations (\ref{h-3})
 and (\ref{h-c3})
  in the form
 \be\label{h-c4}
 \triangle_{\mathrm{BL}} {\psi}=0,~~~\triangle_{\mathrm{BL}} [N_d{\psi}^{7}]=0
 ~~~\to~~~{\psi}=1+\frac{r_g}{r},~~
 [N_d{\psi}^{7}]=1-\frac{r_g}{r},~~N^k=0,
 \ee
 where $r_g$ is the constant of the
 integration given by the boundary conditions.

The question arises: Where is the Hubble evolution in the canonical
GR?

\subsection{Global energy constraint and
 dimension of diffeomorphisms ($3L+1G\neq4L$)}

The Dirac -- ADM approach to the Einstein--Hilbert theory \cite{d}
states that five components $\psi,N_{\rm d},N^k$ are
 treated  as {\it
 potentials} satisfying the Laplace type equations in curved space
  without the initial data, three components
  are excluded by the gauge constraints $\partial_k{\bf e}^k_{(b)}$,
  and only  two rest  transverse gravitons
  are considered as
  independent {\it degrees of freedom}
  satisfying the d'Alambert type equations with the
  initial data.
  This Dirac -- ADM classification is not compatible with both
  the cosmological observations including the last CMB data and
the group of general coordinate transformations
 that conserves
  a family  of constant coordinate time hypersurfaces $x^0=\rm{const}$.
 The group of these transformations, known as
  {\it
kinemetric} subgroup \cite{vlad}, contains only homogeneous
 reparameterizations of the coordinate evolution parameter $(x^0)$ and three local transformations of the spatial coordinates:
  \bea\label{1-8}
  \left[\begin{array}{c}x^0\\x^i\end{array}\right]\to \left[\begin{array}{c}\widetilde{x}^0(x^0)\\\widetilde{x}^i(x^0,{x}^i)\end{array}\right]
 \eea
 This means that dimension of the kinemetric subgroup of
 diffeomorphisms (three local functions and one global one)
 does not coincide with the dimension of the
  constraints in the canonical approach to the classical theory of
  gravitation
  that remove four local variables (the law $3L+1G\neq4L$).

  The kinemetric subgroup (\ref{1-8}) essentially
  simplifies the solution of the energy constraint (\ref{h-3}),
  if the homogeneous variable is extracted from the
  the determinant
  \be\label{d-t1}
  \vh_0\psi^2(x^0,x^k)=\vh(x^0)\widetilde{\psi}^2(x^0,x^k)
  \ee
  with the additional constraints
 \be\label{3-20}
 \int d^3x \log\widetilde{\psi}=\int d^3x \left[\log{\psi}
 -\left\langle{ \log{\psi}}\right\rangle\right]\equiv 0,~~~\langle\log{\psi}\rangle\equiv\frac{1}{V_0}\int
 d^3x\log{\psi},
 \ee
 where $V_0=\int d^3x  < \infty$ is the finite
 Lichnerowicz volume.

According to the definition of
 all measurable quantities  as diffeo-invariants \cite{Dir}, in
 finite space-time the non diffeo-invariant quantity (\ref{1-8})
 $(x^0)$ is not measurable.
  Wheeler and DeWitt \cite{WDW} draw attention that in this case
   evolution of a universe in GR 
  lies in full analogy with a relativistic particle
  given by the action
 \be\label{3-25}
 \widetilde{S}_{\rm SR}[X^0|X^k]\!=\!-\frac{m}{2}
 \int d\tau~ \frac{1}{e_p}\left[\left(\frac{dX^0}{d\tau}\right)^2
 -\left(\frac{dX^k}{d\tau}\right)^2+e^2_p\right]\!=\!\int\! d\tau \left[
 -P_\mu \frac{dX^\mu}{d\tau}+\frac{e_p}{2m}(P^2_\mu-m^2)\right]
 \ee
   in the Minkowski space
  of events $[X^0|X^k]$ and the interval $ds=e_pd\tau$,
  because both the actions (\ref{3-25}) in SR  and (\ref{h-1}) in GR
 are  invariant with respect to reparametrizations of the
 coordinate evolution parameters
 $\tau\to \widetilde{\tau}=\widetilde{\tau}(\tau)$ and
 $x^0\to \widetilde{x}^0=\widetilde{x}^0(x^0)$, respectively.
 In any  relativistic theory given by an action and
 a geometrical interval \cite{H}
 there are two diffeo-invariant
 time-like parameters:
 the diffeo-invariant geometrical proper time interval (g-time)
 $e_pd\tau=ds$ and the one of dynamical variables $X^0$ in the
 {\it space of events} $[X^0|X^k]$ (d-time).

 Therefore, one should points out in the finite volume GR
 the homogeneous variable $\vh(x^0)$
 as the evolution parameter (d-time) in the field space of events
 $[\vh|\widetilde{F}]$
 and diffeo-invariant time coordinate $e_u dx^0=d\zeta$
 (g-time), where $e_u[\widetilde{N}_{\rm d}]$ as functional of
 $\widetilde{N}_{\rm d}$ can be defined as the spacial averaging
 \be\label{3-21-2}
 \frac{1}{e_u[\widetilde{N}_{\rm d}]}=\frac{1}{V_0}\int \frac{d^3x}{\widetilde{N}_{\rm d}}
 \equiv\langle\widetilde{N}_{\rm d}^{-1} \rangle.
 \ee
 This definition is consistent with
 action of GR obtained after the extraction of the d-time
 (\ref{d-t1})
 \cite{bpzz,bard,242}
 \be\label{3-21}
 S[\vh_0|F]=\widetilde{S}_u[\vh|\widetilde{F}]-
 V_0\int dx^0 \frac{1}{e_u}\left(\frac{d\vh}{dx^0}\right)^2=\int dx^0 L;
\ee
 where  $\widetilde{S}[\varphi|\widetilde{F}]$
  is the action (\ref{1-1})  in
 terms of metrics ${\widetilde{g}}$, where  $\vh_0$ is replaced by
 the running scale
 $
 \vh(x^0)=\vh_0a(x^0)
 $ 
  of all masses  of the
 matter fields.  The action (\ref{3-21}) leads to
 the energy constraints
 \be\label{3-22}
 \frac{\delta S[\vh_0]}{\delta
 \widetilde{N}_{\rm d}}=-{T}_{\rm d}=
 \frac{(\partial_0\varphi)^2}{\widetilde{N}_{\rm d}^2}-
 \widetilde{T}_{\rm d}=0,~~ \widetilde{T}_{\rm d}\equiv-
 \dfrac{\delta \widetilde{S}[\vh]}{\delta  \widetilde{N}_{\rm
 d}}\geq 0
 \ee

\section{{\it Canonical} cosmic evolution in the field space of events}

\subsection{The Wheeler -- DeWitt
 universe -- particle correspondence}

\begin{table}[hpb!]
\centering
\caption{The Universe-particle correspondence \cite{bpzz,242}.}
\begin{tabular}{|c|c|c|c|}
\hline $\mathrm{N}^{\underline{\mathrm{o}}}$ & concepts &
universe& particle\\[.1cm]
\hline 1.&reparametrizations & $x^0 \to
 \widetilde{x}^0=\widetilde{x}^0(x^0)$
& $\tau \to
 \widetilde{\tau }=\widetilde{\tau }(\tau )$\\[.1cm]
\hline 2.& evolution parameter & $\vh(x^0)=\vh_0a(x^0)$ & $X_0(\tau)$\\[.1cm]
\hline
 3.& space of events & $\vh~|~\widetilde{F}$ & $X_0~|~X_k$\\[.1cm]
\hline
 4.& geometric time  & $d\zeta=dx^0e_u$ & $ds=d\tau e_p$\\[.1cm]
\hline 5.& the arrow of time &
$\zeta_{(\pm)}=\pm\int^{\vh_0}_{\vh_I}
{d\vh}~{\langle{(\widetilde{T}_{\rm d}})^{-1/2}\rangle}\geq 0$ &
$s_{\pm}
=\pm\frac{m}{E}[X_0^0-X^0_I]\geq 0$\\[.1cm]
\hline
 6.& energy constraint  & $P^2_\vh-E^2_\vh=0$ & $ P^2_0-E^2_0=0$\\[.1cm]
\hline 7.& energy of events & $P_\vh=\pm E_\vh=\pm 2 \int
d^3x(\widetilde{T}_{\rm d})^{1/2}$ & $P_0=\pm
E_0=\pm\sqrt{m^2c^4+|\vec p|^2}$\\[.1cm]
\hline 8.& wave equation & $[\hat P^2_\vh-E^2_\vh]\Psi_{\rm WDW}=0$
&
 $[\hat P^2_0-E^2_0]\Psi_{\rm KG}=0$\\[.1cm]
\hline
 9.&  secondary quantization & $\Psi_{\rm
 WDW}=[{1}/{\sqrt{2E_\vh}}][A^++A^-]$ &
 $\Psi_{\rm
KG}=[{1}/{\sqrt{2E_0}}][a^++a^-]$\\[.1cm]
 \hline
 10. & Bogoliubov transformation & $A^+=\alpha
 B^+\!+\!\beta^*B^-$ &
 $ a^+=\alpha
 b^+\!+\!\beta^*b^-$\\[.1cm]
 \hline
 11.&  the stable vacuum state & $B^-|0>=0$ &
 $b^-|0>=0$\\[.1cm]
\hline 12.& creation from vacuum & $N_{\rm
universe}=<0|A^+A^-|0>\not =0$ &
 $N_{\rm particle}=<0|a^+a^-|0> \not =0$\\[.1cm]
\hline
\end{tabular}
\end{table}

  According to the Wheeler -- DeWitt \cite{WDW} there is the
 universe -- particle correspondence  given in the Table 1 \cite{bpzz,242}.
 This universe-particle correspondence
 rejects the Hilbert {\it Foundations
 of relativistic physics} of 1915 \cite{H} that include
 also the geometric interval (Table 1.4)
 and the group of diffeomorphisms (Table 1.1), in contrast to the
 classical physics based  only on an action and the group of
 the data transformations.
 The group of diffeomorphisms (Table 1.1) leads to the energy
 constraint (Table 1.6). Resolution of the energy constraint
  gives the Hubble type relation (Table 1.5) between d-time (Table 1.3)
   and g-time (Table 1.4) and determines the energy of events (Table 1.7)
   that can take  positive and negative values. In aim to remove
   the negative value, one can use the rich experience of QFT, i.e.
  the primary quantization (Table 1.8) and the secondary one (Table 1.9).
  This quantization procedure leads immediately to
   creation from stable Bogoliubov vacuum state (Table 1.12) of both quasiuniverses and
   quasiparticles (Table 1.11) obtained by the Bogoliubov transformation (Table 1.10)
\cite{61,69}.
 This QFT experience illustrates  the possibility to  solve
 the problems of the quantum origin of all matter fields
  in the Early Universe,  its evolution,
  and the present-day energy budget \cite{bpzz,origin,114:a}.
  In order to use this possibility, one should impose a set of
 requirements on the cosmic motion in the field space of events
 that follow from the general principles of QFT.

\subsection{CMB requirements to the {\it canonical}  cosmological perturbation
theory}

    The  QFT experience  supposes that the action
 (\ref{3-21})  can be represented in the {\it canonical}
 Hamiltonian form like (\ref{h-1})
 \be\label{1-36}
 S_{\rm canonical}[\vh_0|F]=\int dx^0\left\{-P_{\vh}\partial_0\vh+
 e_u[\widetilde{N}_{\rm d}]\frac{P^2_\vh}{4V_0}+\int d^3x
 \left[\sum\limits_{\widetilde{F}
 } P_{\widetilde{F}}\partial_0\widetilde{F}
 +{\cal C}-\widetilde{N}_d \widetilde{T}_{\rm d}^0\right]\right\}.
 \ee
 However, the acceptable cosmological perturbation theory \cite{Lif,kodama,bard} \emph{is
 not compatible}
with the Hamiltonian  formulation  (\ref{1-36}), because of after
the separation of cosmological scale factor in the common accept
cosmological perturbation theory the number of variables does not
coincide with the number of variables in GR \cite{242}.

In this case, the energy constraint (\ref{3-22}) takes the form of
\emph{the Friedmann equation}
 \be\label{6-1-113ec}
 \left[\frac{d\varphi}{d\zeta}\right]^2\equiv\vh'^2=
 {\left\langle(\widetilde{T}_{\rm d})^{1/2}\right\rangle}^2 ,
 \ee
 and the algebraic equation for
the diffeo-invariant lapse function
 \be\label{3-29}
  N_{\rm inv}=\langle(\widetilde{N}_{\rm d})^{-1} \rangle
 \widetilde{N}_d=
 \left\langle (\widetilde{T}_{\rm d})^{1/2}\right\rangle
 ({{\widetilde{T}_{\rm d}}})^{-1/2}.
 \ee
We see that the energy constraint (\ref{3-22}) removes
  only one  global momentum $P_\vh$
 in accord to the dimension
  of the kinemetric diffeomorphisms (\ref{1-8})
  that is consistent with the second Noether theorem.

 One can find the
 evolution of all field variables $F(\vh,x^i)$  with respect to
 $\vh$ by variation of the ``reduced'' action
 \be\label{2ha2} S[\vh_0]{}_{{}_{{P_\vh=\pm E_\vh}}} =
 \int\limits_{\vh_I}^{\vh_0}d\widetilde{\vh} \left\{\int d^3x
 \left[\sum\limits_{  F}P_{  F}\partial_\vh F
 +\bar{\cal C}\mp2\sqrt{\widetilde{T}_{\rm d}(\widetilde{\vh})}\right]\right\},
\ee
 obtained as
   values of the Hamiltonian form of the initial action   (\ref{1-36})
 onto the energy constraints (\ref{6-1-113ec}), where $\bar{\cal C}={\cal
 C}/\partial_0\widetilde{\vh}$ \cite{pp}.

 The energy constraints (\ref{3-22}) and the Hamiltonian reduction  (\ref{2ha2})
 lead to the  definite
 {\it  canonical   rules} of the Universe evolution in the field space of
 events $[\vh|\widetilde{F}]$.

\begin{description}
 \item[Rule 1: Causality Principle in the WDW space] $\dfrac{d\vh_I}{d\vh_0} =0${\textit{
  follows from the Hamiltonian reduction
 (\ref{2ha2}) that gives us the solution of the Cauchy problem
 and means that initial data do not
 depend on the Planck value.
 }}
 \item[Rule 2: Positive
 Energy Postulate] \emph{follows from
 the energy constraint} (\ref{3-22})
  $\dfrac{(\partial_0\varphi)^2}{\widetilde{N}_{\rm d}^2}=
 \widetilde{T}_{\rm d}${
 \begin{equation}\label{rule-I}
 \widetilde{T}_{\rm d}=-\dfrac{16}{\vh^2}
p^2_{\widetilde{\psi}}+...\geq 0 ~~~~~~~~ \to~~~~~~~~
 p_\psi=-\frac{4\vh^2}{3{\widetilde{\psi}^6N_{\rm
 inv}}}[\partial_j(\widetilde{\psi}^6{\cal
 N}^j)-(\widetilde{\psi}^6)']=0,.
 \end{equation}
 \emph{where
 $\widetilde{T}_{\rm d}$ is given by equations of the type of (\ref{h-3})
  and (\ref{h-5}) and $ ({\cal N}^j=N^j\langle
 N^{-1}_d\rangle\not =0)$.}
 }
 \item[Rule 3: Vacuum Postulate]{\textit{ $B^-|0>=0$
 restricts  the  Universe motion in the field space of events
 \begin{eqnarray}
 P_\vh\geq 0~~\mathrm{for}~~\vh_I\leq\vh_0\\\nonumber P_\vh\leq
 0~~\mathrm{for}~~\vh_I\geq\vh_0.
 \end{eqnarray}
 }}
 \item[Rule 4: Lapse Function ]{\textit{$N_{\rm inv}>0$
 follows from
 the nonzero energy density $\widetilde{T}_{\rm d}\not =0$.
 }}
\end{description}

The Rule 1  is not compatible with the Planck
 epoch in the beginning of the Universe $\dfrac{d\vh_I}{d\vh_0}\not =0$.

 The Rule 2 is not compatible with dynamic evolution of the local scalar
 component $\widetilde{\psi}^2=\psi^2/a$.

 The Rule 3 leads to the arrow of the geometric time.

 The Rule 4 forbids any penetration into a internal region of black
 hole because this penetration is accompanied the change of
 a sign of the lapse function.

\begin{table}[hpb!]
\centering \caption{The canonical cosmological perturbation
theory\cite{bpzz,242} versus the standard one
\cite{Lif,kodama,bard}.}
\begin{tabular}{|c|c|c|c|}
\hline $\mathrm{N}^{\underline{\mathrm{o}}}$ & concepts &
Canonical Cos. P.Th.& Standard Cos. P.Th\\
\hline 1. & Number of variables & It is equal to the GR one &
It is not equal to the GR one\\
\hline 2.&frame &  CMB frame  & frame free\\
\hline
 3.& Planck's epoch &It is  at present-day & It is at Early Universe\\
\hline
 4.& geometric time  & diffeo-invariant $d\zeta=dx^0e_u$ & diffeo-variant $\eta$\\
\hline 5.& the arrow of time &
$\zeta_{(\pm)}=\pm\int^{\vh_0}_{\vh_I}
{d\vh}~{\langle{(\widetilde{T}_{\rm d}})^{-1/2}\rangle}\geq 0$ &
It is not\\
\hline 6.& energy of events & $P_\vh=\pm E_\vh=\pm 2 \int
d^3x(\widetilde{T}_{\rm d})^{1/2}$ & It is not\\
\hline
 7.&  Kinetic perturbations &  $p_\psi=0$ &
 $p_\psi\not=0$\\
 \hline
 8. & Shift vector $N^k$ & $N^k\not =0$&
  $N^k =0$\\
 \hline
 9.&  vacuum postulate & vacuum exist &
 vacuum not exist\\
 \hline
 10.& Potential perturbations & $T_{d}(\psi)=T_{d}(\psi=a^{1/2})+\Delta T_{d}$ &
$T_{d}(\psi)=T_{d}(\psi=a^{1/2})$\\ \hline
\end{tabular}
\end{table}

 Thus, the explanation of the quantum origin of the Universe
 in GR (Table 2.1.) and its matter in the framework
 of the canonical GR  is not compatible with
  the  frame free cosmology (Table 2.2.),
 the Planck's initial data at the Early Universe (Table 2.3.)
  and the scalar component dynamics (Table 2.8.)
 considered as the basis  of the Inflationary model \cite{linde}.

  First of all one should check
 the correspondence of the canonical GR
  with both the QFT in the flat space-time and
  the classical Newton theory.

\subsection{Test I. The QFT limits and SN data}

 \begin{figure}[t]
\centering\includegraphics[scale=0.68]{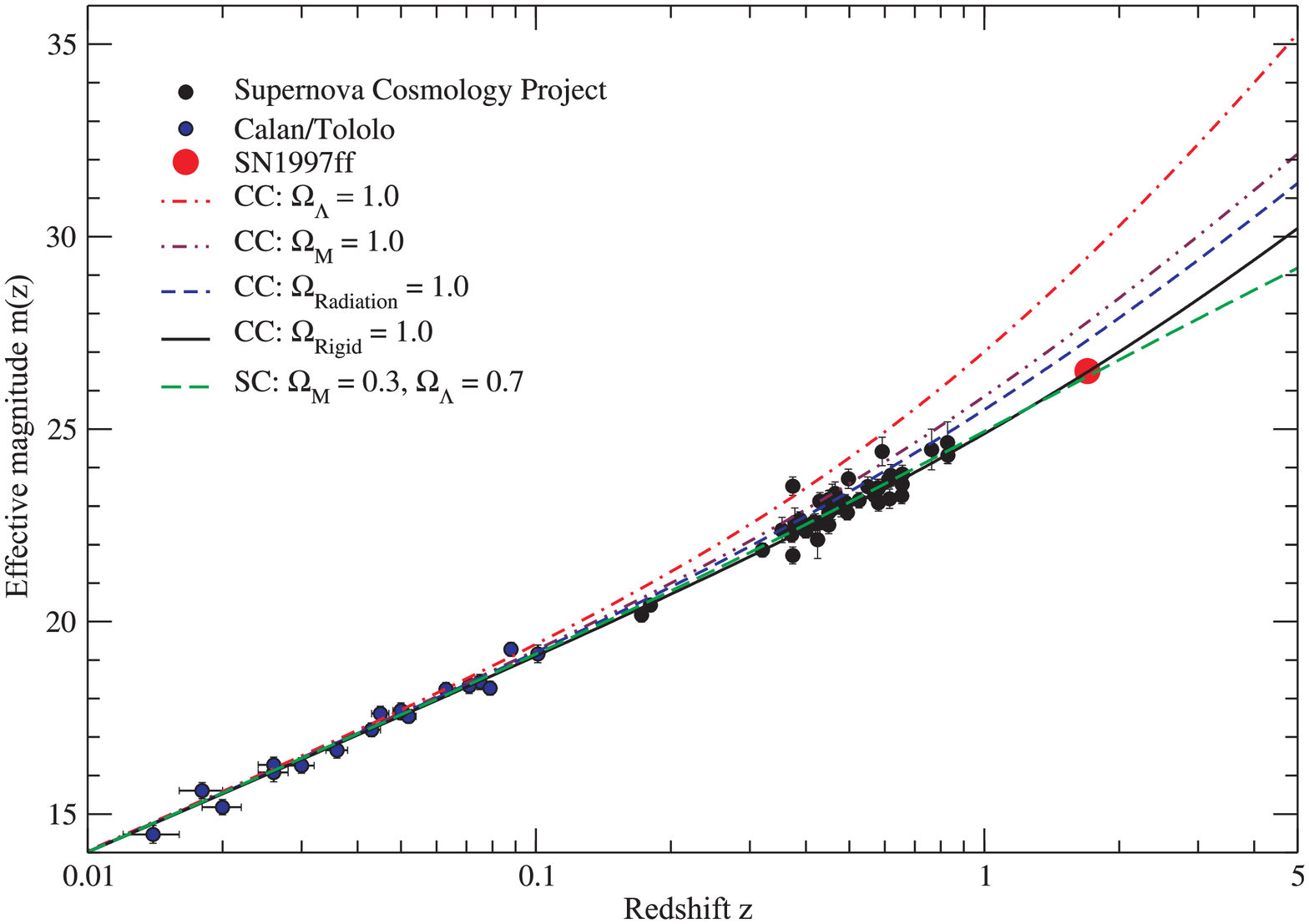} \caption{The Hubble
diagram in cases of the {\it absolute} units of standard cosmology
(SC) and the {\it relative} units of conformal cosmology (CC)
\cite{039a, Danilo, zakhy}.
 The points include 42 high-redshift Type Ia
 supernovae \cite{snov} and the reported
 farthest supernova SN1997ff \cite{SN}. The best
fit to these data  requires a cosmological constant
$\Omega_{\Lambda}=0.7$, $\Omega_{\rm CDM}=0.3$ in the case of SC,
whereas in CC
 these data are consistent with  the dominance of the rigid (stiff)
 state. The Hubble Scope Space Telescope team analyzed 186 SNe Ia \cite{riess1} to test the CC
 \cite{zakhy}.
 \label{fig1}
 }
\end{figure}

 The correspondence principle \cite{pp}
 as the low-energy
 expansion of the {\it``reduced action''} (\ref{2ha2}) over the
 field density ${T}_{{\rm s}}$
 \be
 2d\vh \sqrt{T_0^0}= 2d\vh
 \sqrt{\rho_{0}(\vh)+{T}_{{\rm s}}}
 =
 d\vh
 \left[2\sqrt{\rho_0(\vh)}+
 \frac{{T}_{{\rm s}}}{\sqrt{\rho_0(\vh)}}\right]+...
 \ee
 gives the following sum:
 \be
 S^{(+)}|_{\rm
 constraint}= S^{(+)}_{\rm cosmic}+S^{(+)}_{\rm
 field}+\ldots,
 \ee where
\be
 S^{(+)}_{\rm cosmic}[\varphi_I|\varphi_0]= -
 2V_0\int\limits_{\vh_I}^{\vh_0}\!
 d\vh\!\sqrt{\rho_0(\vh)}
 \ee is the reduced  cosmological action (\ref{2ha2}),
 and
 \be\label{12h5} S^{(+)}_{\rm field}=
 \int\limits_{\eta_I}^{\eta_0} d\eta\int\limits_{V_0}^{} d^3x
 \left[\sum\limits_{ F}P_{ F}\partial_\eta F
 +\bar{{\cal C}}-{T}_{{\rm s}}\right]
 \ee
 is the standard field action
 in terms of the conformal time:
 $d\eta=\dfrac{d\vh}{\sqrt{\rho_0(\vh)}}$,
 in the conformal flat space--time with running masses
 $m(\eta)=a(\eta)m_0$.

 This expansion shows that the Hamiltonian approach to the General
 Theory of Relativity
 in terms of the Lichnerowicz scale-invariant variables
 (\ref{adm-2}) identifies the ``conformal quantities''
  with the observable ones including the conformal time $d\eta$,
  instead of $dt=a(\eta)d\eta$, the coordinate
 distance $r$, instead of the Friedmann one $R=a(\eta)r$, and \emph{the conformal
 temperature $T_c=Ta(\eta)$, instead of the standard one $T$}.
 Therefore,
 the scale-invariant variables  distinguish the conformal
 cosmology (CC) \cite{039, Narlikar},
  instead of the standard cosmology (SC).
 In this case,
 the
  red shift of the wave lengths of the photons
  emitted at the time $\eta_0-r$ by atoms on a cosmic object in the comparison
  with the Earth ones emitted at emitted at the time $\eta_0$,
  where $r$ is the distance between the Earth and the object:
\be \frac{\lambda_0}{\lambda_{\rm
cosmic}(\eta_0-r)}=\frac{a(\eta_0-r)}{a (\eta_0)}\equiv a(\eta_0-r)
=\frac{1}{1+z}. \ee
 This red shift can be  explained by the running masses
 $m=a(\eta)m_0$ in action (\ref{12h5}). In this case, the
 Schr\"odinger wave equation \bea\label{schL}
 \left[\frac{\hat p^2_r}{2 a(\eta)m_0}-\frac{\alpha}{r}\right]\Psi_L(\eta,r)=
 \frac{d}{id\eta}\Psi_L(\eta,r)
 \eea
 can be converted
 by the substitution $r=\dfrac{R}{a(\eta)}$, $p_r=P_{R}a(\eta)$, $a(\eta)d\eta=dt$,
 $a(\eta)\Psi_L(\eta,r)=\Psi_{0}(t,R)$ into the
 standard Schr\"odinger wave equation with the constant mass
\bea\label{sch0}
 \left[\frac{\hat P^2_R}{2 m_0}-\frac{\alpha}{R}\right]\Psi_0(t,R)=
 \frac{d}{i dt}\Psi_0(t,R).
 \eea
 Returning back to the Lichnerowicz variables $\eta,r$
 we obtain the spectral decomposition of the wave
 function of an atom with the running mass
 \bea\label{1sch0}
 \Psi_L(\eta,r)=\frac{1}{a(\eta)}\sum\limits_{k=1}^{\infty}
 e^{-i\varepsilon_0^{(k)}
 \int\limits_{\eta}^{\eta_0}d\widetilde{\eta}a(\widetilde{\eta})}\Psi^{(k)}_0(a(\eta)r)=\sum\limits_{k=1}^{\infty}\Psi^{(k)}_L(\eta,r).
 \eea
 Where $\varepsilon_0^{(k)}=\alpha^2m_0/k^2$ is a set of eigenvalues
 of the Schr\"odinger wave equation in the Coulomb potential. We got the
 equidistant spectrum $-i(d/d\eta)\Psi^{(k)}_L(\eta,r)=
 \varepsilon_0^{(k)}\Psi^{(k)}_L(\eta,r)
 $ for any wave lengths of cosmic photons
 remembering the size of the atom at the moment of their emission.

 The conformal observable distance  $r$ loses the factor $a$, in
 comparison with the nonconformal one \mbox{$R=ar$}. Therefore, in the
 case of CC, the redshift --
  coordinate-distance relation $d\eta=\dfrac{d\vh}{\sqrt{\rho_0(\vh)}}$
  corresponds to a different
  equation
  of state than in the case of SC \cite{039}.
 The best fit to the data,  including
  Type Ia supernovae~\protect \cite{snov, SN},
 requires a cosmological constant $\Omega_{\Lambda}=0.7$,
$\Omega_{\rm CDM}=0.3$ in the case of the ``scale-variant
quantities`` of standard cosmology. In the case of ``conformal
 quantities'' in CC, the Supernova data \cite{snov, SN} are
consistent with the dominance of the stiff (rigid) state,
$\Omega_{\rm Rigid}\simeq 0.85 \pm 0.15$, $\Omega_{\rm Matter}=0.15
\pm 0.15$ \cite{039, 039a, Danilo}. If $\Omega_{\rm Rigid}=1$, we
have the square root dependence of the scale factor on conformal
time $a(\eta)=\sqrt{1+2H_0(\eta-\eta_0)}$. Just this time dependence
of the scale factor on
 the measurable time (here -- conformal one) is used for description of
 the primordial nucleosynthesis \cite{Danilo, three}.

 This stiff state is formed by a free scalar field
 when $E_\vh=2V_0\sqrt{\rho_0}=\dfrac{Q}{\vh}$. In this case there is an exact
solution of the Bogoliubov equations  of the number of universes
created from a vacuum with the initial data
$\vh(\eta=0)=\vh_I,H(\eta=0)=H_I$ \cite{origin}.

\subsection{Test II: Cosmological creation of CMB and matter}

 These initial data $\vh_I$ and $H_I$ are determined by the
 parameters of matter cosmologically created from the Bogoliubov
 vacuum  at the beginning of a universe $\eta\simeq 0$.

\begin{figure}[hpb!]
\begin{center}
 \includegraphics[scale=0.35,angle=-90]{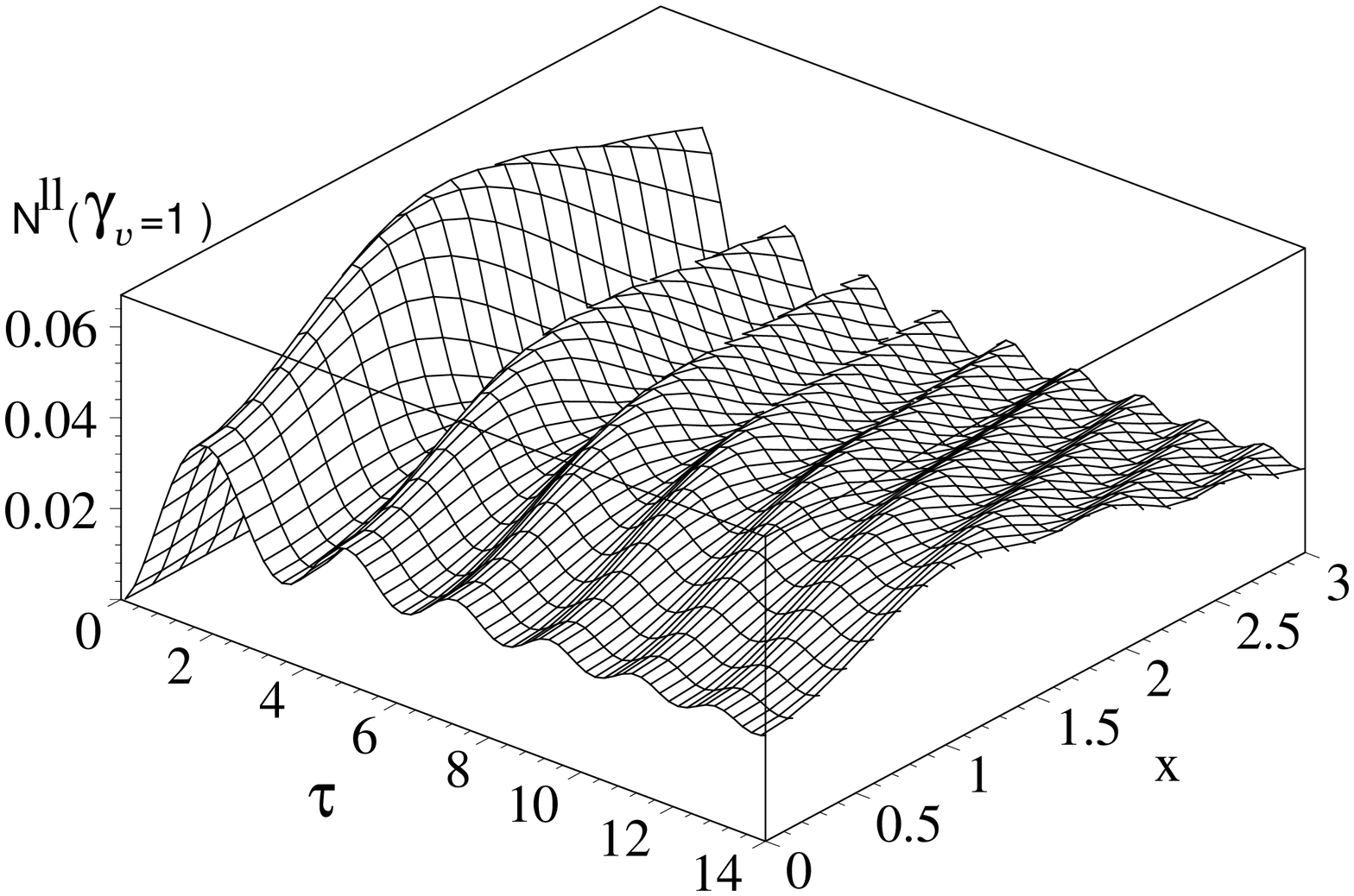}\hspace{-5mm}
 \includegraphics[scale=0.35,angle=-90]{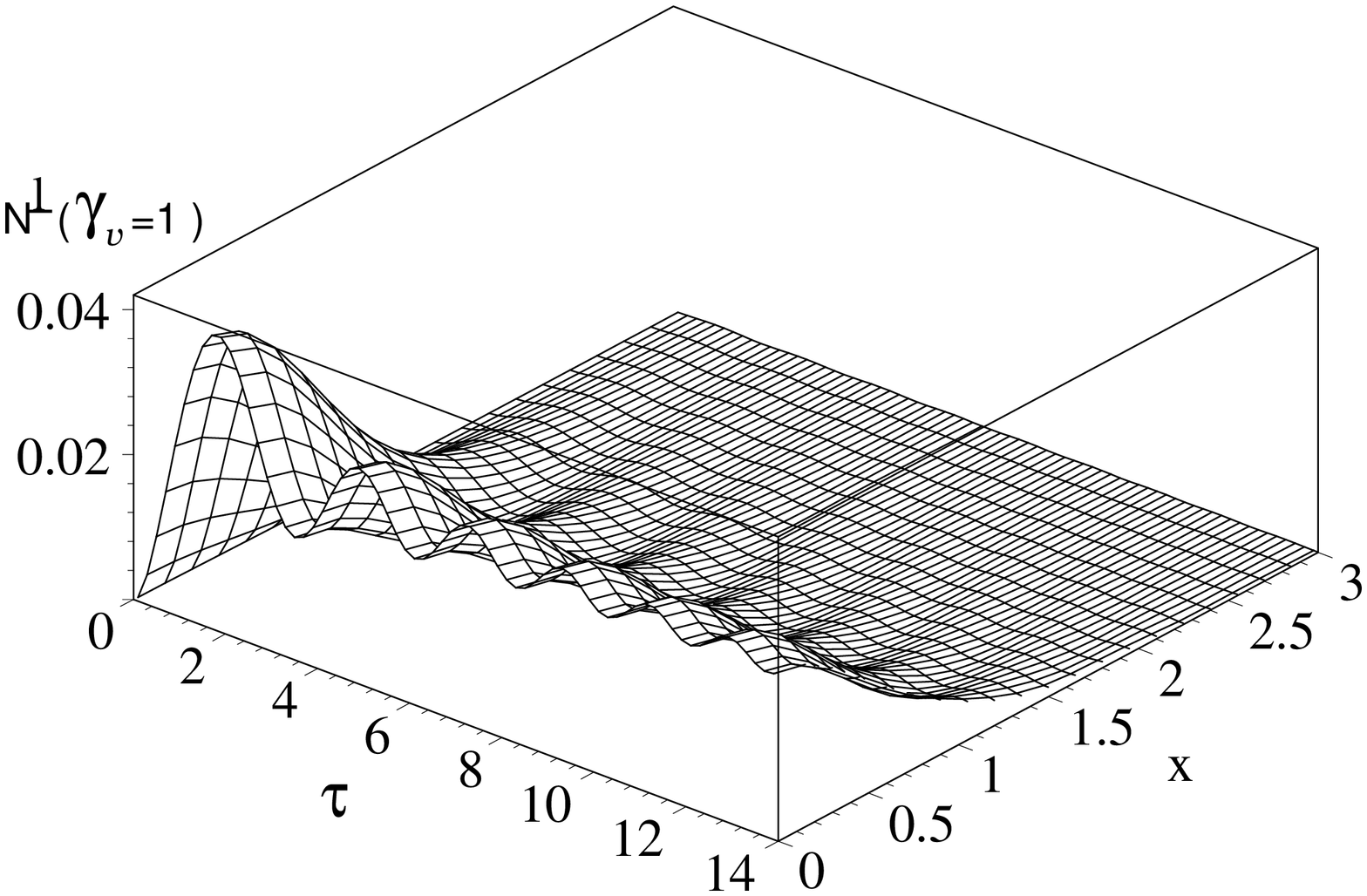}
\parbox{16cm}{  \caption{
Dependence of longitudinal ${N^\|}$ and transverse ${N^\bot }$
components of the distribution function of vector bosons  on time
$\tau=2h_i\eta$ and momentum $(x=q/m_i)$. Their momentum
distributions in  units of the primordial mass $ x = q/M_I $ show
the large contribution of longitudinal bosons. Values of the initial
data $M_I = H_I$  follow from the uncertainty principle and give the
temperature of relativistic bosons $T\sim H_I=(M_0^2H_0)^{1/3}=2.7 K
$ \cite{114:a}
  }}
\end{center}
\end{figure}

 The Standard
 Model (SM) density ${T}_{{\rm s}}$ in action (\ref{12h5})
  shows
 us that W-, Z- vector bosons have maximal probability of this
 cosmological creation
 due to their mass singularity \cite{114:a}. One can introduce the notion of
\emph{a particle in a universe} if the Compton length of a particle
 defined by its inverse mass
 \mbox{$M^{-1}_{\rm I}=(a_{\rm I} M_{\rm W})^{-1}$} is less than the
 universe horizon defined by the inverse Hubble parameter
 $H_{\rm I}^{-1}=a^2_{\rm I} (H_{0})^{-1}$ in the
 stiff state. Equating these quantities $M_{\rm I}=H_{\rm I}$
 one can estimate the initial data of the scale factor
 \mbox{$a_{\rm I}^2=(H_0/M_{\rm W})^{2/3}=10^{-29}$} and the primordial
 Hubble parameter
 \mbox{$H_{\rm I}=10^{29}H_0\sim 1~{\rm mm}^{-1}\sim 3K$}.
 Just at this moment there is  an effect of intensive
  cosmological creation of the vector bosons described in papers
  \cite{114:a,vin} (see Fig. 2);
 in particular, the distribution functions of the longitudinal vector bosons
demonstrate clearly a large contribution of relativistic momenta.
 Their conformal (i.e. observable) temperature $T_c$
 (appearing as  a consequence of
 collision and scattering of these bosons) can be estimated
from the equation in the kinetic theory for the time of
establishment of this temperature \mbox{$\eta^{-1}_{relaxation}\sim
n(T_c)\times\sigma\sim H$}, where $n(T_c)\sim T_c^3$ and $\sigma
\sim 1/M^2$ is the cross-section. This kinetic equation and values
of the initial data $M_{\rm I} = H_{\rm I}$ give the temperature of
relativistic bosons \be\label{t}
 T_c\sim (M_{\rm I}^2H_{\rm I})^{1/3}=(M_0^2H_0)^{1/3}\sim 3 K
\ee as a conserved number of cosmic evolution compatible with the
Supernova data \cite{039,snov,SN}.
 We can see that
this  value is surprisingly close to the observed temperature of the
CMB radiation
 $ T_c=T_{\rm CMB}= 2.73~{\rm K}$.

 The primordial mesons before
 their decays polarize the Dirac fermion vacuum
 (\emph{as the origin of axial anomaly} \cite{riv,ilieva,gip,j})
 and give the
 baryon asymmetry frozen by the CP -- violation.
The
 value of the baryon--antibaryon asymmetry
of the universe following from this axial anomaly was estimated in
paper \cite{114:a} in terms of the coupling constant of the
superweak-interaction
 \be\label{a1}
 n_b/n_\gamma\sim X_{CP}= 10^{-9}.
 \ee
The boson life-times     $\tau_W=2H_I\eta_W\simeq
\left({2}/{\alpha_W}\right)^{2/3}\simeq 16,~ \tau_Z\sim
2^{2/3}\tau_W\sim 25
 $ determine the present-day visible
baryon density
\be\label{b}\Omega_b\sim\alpha_W=\dfrac{\alpha_{QED}}{\sin^2\theta_W}\sim0.03.\ee
All these results (\ref{t}) -- (\ref{b})
 testify to that all  visible matter can be a product of
 decays of primordial bosons, and the observational data on CMB
 radiation
 can reflect  parameters of the primordial bosons, but not the
 matter at the time of \emph{recombination}. In particular,
 the length of  the semi-circle on the surface of  the last emission of
photons at the life-time
  of W-bosons
  in terms of the length of an emitter
 (i.e.
 $M^{-1}_W(\eta_L)=(\alpha_W/2)^{1/3}(T_c)^{-1}$) is
 $\pi \cdot 2/\alpha_W$.
 It is close to $l_{min}\sim  210 $ of CMB radiation,
 whereas $(\bigtriangleup T/T)$ is proportional to the inverse number of
emitters~
 $(\alpha_W)^3 \sim    10^{-5}$.

 The temperature history of the expanding universe
 copied in the ``conformal quantities'' looks like the
 history of evolution of masses of elementary particles in the cold
 universe with the constant conformal temperature $T_c=a(\eta)T=2.73~ {\rm K}$
 of the Cosmic Microwave Background radiation.

 Equations of the
 vector bosons
 in SM are very close  to
 the  equations of the $\Lambda$CDM model
  with  the inflationary scenario used
  for description  of  the CMB ``power primordial spectrum''.

 \subsection{Test III: The Newton potential and the Large-scale structure}
 The cosmological generalization of the static potential in terms
 of the Lichnerowicz variables is as follows
\bea\label{adm-2}
 \omega^{(L)}_{(0)}&=&\widetilde{\psi}^4N_{\rm int}d\zeta,~~~~~
 \omega^{(L)}_{(b)}={\bf e}_{(b)k}[dx^k +{\cal N}^kd\zeta],
 \\
\label{La-2}
 \widetilde{T}_{\rm d}&=& \widetilde{\psi}^{7}\hat \triangle
 \widetilde{\psi}+
  \sum\limits_{I} \widetilde{\psi}^I a^{\frac{I}{2}-2}{\cal T}_I,
  ~~~~~~{\cal T}_I\equiv\langle{\cal T}_I\rangle+\overline{{\cal T}_I}
 \eea
and is determined by the Eqs.(\ref{3-29}), (\ref{rule-I}), and
$\overline{T}_\psi=0$,
    instead of the
   the infinite volume GR ones (\ref{h-3}) and (\ref{h-c3}).
\be\label{n-1}
 T_{\rm d}=T_{\psi}=p_\psi=0 ~~~\Longrightarrow~~N_{\rm int}=
 \frac{\left\langle\sqrt{\widetilde{T}_{\rm d}}\right\rangle}{\sqrt{\widetilde{T}_{\rm d}}}
 , ~~~~T_{\psi}=p_\psi=0
\ee

 The choice of the L-variables  (\ref{adm-2}) and (\ref{La-2})
 determines $\widetilde{\psi}$ and $N_{\rm inv}$ in the form of a sum
  \bea\label{12-17}
 \widetilde{\psi}&=&1+\frac{1}{2}\int d^3y\left[D_{(+)}(x,y)
\overline{T}_{(+)}^{(\mu)}(y)+
 D_{(-)}(x,y) \overline{T}^{(\mu)}_{(-)}(y)\right],\\\label{12-18}
 N_{\rm inv}\widetilde{\psi}^7&=&1-\frac{1}{2}\int d^3y\left[D_{(+)}(x,y)
\overline{T}^{(\nu)}_{(+)}(y)+
 D_{(-)}(x,y) \overline{T}^{(\nu)}_{(-)}(y)\right],
  \eea
 where
 \be\label{1cur1}\overline{T}^{(\mu)}_{(\pm)}=\overline{{\cal T}_{(0)}}\mp7\beta
  [7\overline{{\cal T}_{(0)}}-\overline{{\cal T}_{(1)}}],
 ~~~~~~~
 \overline{T}^{(\nu)}_{(\pm)}=[7\overline{{\cal T}_{(0)}}-\overline{{\cal T}_{(1)}}]
 \pm(14\beta)^{-1}\overline{{\cal T}_{(0)}}
 \ee
 are the local currents, $D_{(\pm)}(x,y)$ are the Green functions satisfying
 the equations
 \bea\label{2-19}
 [\pm \hat m^2_{(\pm)}-\hat \triangle
 ]D_{(\pm)}(x,y)=\delta^3(x-y),
 \eea
 where $\hat m^2_{(\pm)}= 14 (\beta\pm 1)\langle {\cal T}_{(0)}\rangle \mp
\langle {\cal T}_{(1)}\rangle$, and $\beta$ is given by the equation
 \be\label{beta}
 \beta=\sqrt{1+\frac{\langle{\cal T}_{(2)}\rangle-14\langle{\cal T}_{(1)}\rangle}{98\langle {\cal T}_{(0)}\rangle}}.
 \ee.

  In the case of point mass distribution in a finite volume $V_0$ with the
zero pressure
  and  the  density
  $\overline{{\cal T}_{(0)}}(x)=\dfrac{\overline{{\cal T}_{(1)}}(x)}{6}
  \equiv  M\left[\delta^3(x-y)-\dfrac{1}{V_0}\right]$,
 solutions   (\ref{12-17}),  (\ref{12-18}) take
 the following  form
 \bea\label{12-21}
  \widetilde{\psi}&=1+
  \dfrac{r_{g}}{4r}\left[{\gamma_1}e^{-m_{(+)}(z)
 r}+ (1-\gamma_1)\cos{m_{(-)}(z)
 r}\right],\\\label{12-22}
 N_{\rm inv}\widetilde{\psi}^{7}&=1-
 \dfrac{r_{g}}{4r}\left[(1-\gamma_2)e^{-m_{(+)}(z)
 r}+ {\gamma_2}\cos{m_{(-)}(z)
 r}\right],
 \eea
 where
 $
  {\gamma_1}=\dfrac{1+7\beta}{2},~~~
 {\gamma_2}=\dfrac{14\beta-1}{28\beta},~~
 r_{g}=\dfrac{3M}{4\pi\vh^2},~~
 r=|x-y|.
 $
 These solutions  have spatial oscillations and the
nonzero shift of the coordinate
  origin that  leads to the large scale distribution of the matter  \cite{bpzz}.

In the infinite volume limit $\langle {\cal T}_{(n)}\rangle=0,~a=1$
 solutions (\ref{12-21}) and  (\ref{12-22}) coincide with
 the isotropic version of  the Schwarzschild solutions:
 $\widetilde{\psi}=1+\dfrac{r_g}{4r}$,~
 ${N_{\rm inv}}\widetilde{\psi}^{7}=1-\dfrac{r_g}{4r}$,~$N^k=0$.
 However, the Black Hole generalization is forbidden by the energy
 constraint (\ref{3-29}) $N_{\rm d}\geq 0$.

\begin{figure}[t]
 \begin{center}
 \includegraphics[scale=0.5]{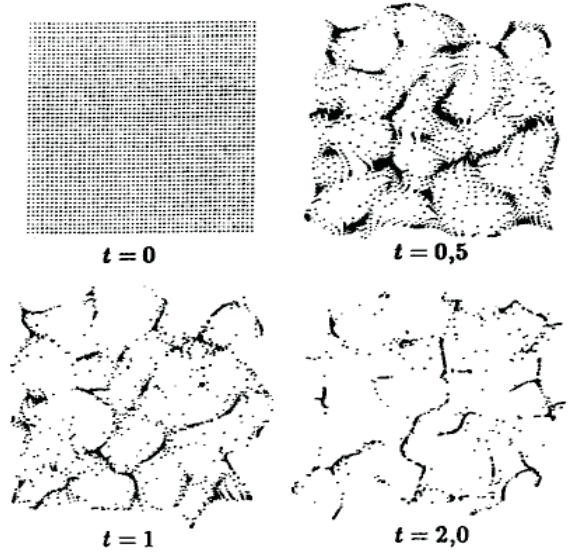}
\caption{
 \label{fig2}
The
 diffusion of a system of particles moving in the space
 $ds^2=d\eta^2-(dx^i+N^id\eta)^2$ with periodic shift vector $N^i$
 and zero momenta could be understood from analysis of O.D.E.
 $dx^i/d\eta=N^i$ considered in the two-dimensional case in the book \cite{kl},
 if we substitute $t= m_{(-)}\eta$ and
 \mbox{$N^i \sim \frac{x^i}{r}\sin m_{(-)}r$} in the equations,
 where $ m_{(-)}$ is defined by Eq.~(\ref{2-19}).
  }
\end{center}
\end{figure}

 In the contrast to standard cosmological perturbation theory \cite{Lif,kodama,bard}
   the diffeo-invariant version of the perturbation theory
 do not contain time derivatives that are responsible for
the CMB ``primordial power spectrum'' in the inflationary model
\cite{bard}. However, the diffeo-invariant version of the Dirac
Hamiltonian approach to GR gives another possibility to explain the
CMB radiation "spectrum" and other topical problems of cosmology by
cosmological creation of the vector bosons considered above. The
equations describing the longitudinal vector bosons
 in SM, in this case, are close to
 the equations that follow from
 the Lifshits perturbation theory and are  used, in  the inflationary model, for
 description of the ``power primordial spectrum'' of the CMB radiation.

 The next differences are a nonzero shift vector and  spatial oscillations of
 the scalar potentials determined by $\hat m^2_{(-)}$ (see Fig. \ref{fig2}).
 In the scale-invariant version of cosmology, \cite{039} the
  SN data dominance of stiff state $\Omega_{\rm Stiff}\sim 1$ determines the parameter
  of spatial oscillations
  \begin{equation}\hat m^2_{(-)}=\dfrac{6}{7}H_0^2[\Omega_{\rm R}(z+1)^2+\dfrac{9}{2}\Omega_{\rm
  Mass}(z+1)].\end{equation}
The redshifts  in the recombination
  epoch $z_r\sim 1100$ and the clustering parameter \cite{kl}
\begin{equation}
r_{\rm clustering}=\dfrac{\pi}{\hat m_{(-)} }\sim \dfrac{\pi}{
 H_0\Omega_R^{1/2} (1+z_r)} \sim 130\, {\rm Mpc}
 \end{equation}
  recently
 discovered in the researches of a large scale periodicity in redshift
 distribution \cite{a1,a2}
 lead to a reasonable value of the radiation-type density
  $10^{-4}<\Omega_R\sim 3\cdot 10^{-3}<5\cdot 10^{-2}$ at the time of this
  epoch.

\section{Geometrization of the Higgs particles
in the unified theory }

We listed a set of arguments in favor that the cosmological problems
and CMB anisotropy can be explained in the framework of the
canonical GR, if we accept the relative units and dilaton GR, for
which the Hilbert action
  (\ref{1-1}) formally coincides with the action
  the dilaton gravitation (DG) \cite{039a}
\be\label{dg1}
  S_{DG}[w,g]=-\int d^4x\frac{\sqrt{-\hat g^w}}{6}~R(\hat g^w)\equiv
  -\int d^4x\left[\frac{\sqrt{- g}w^2}{6}~R( g)-w
  \partial_\mu(\sqrt{- g}\partial_\nu w g^{\mu\nu})\right],
 \ee
 where $\hat g^w=w^2g$ and
   $w$ is the dilaton scalar field. This action is invariant with
 respect to the scale transformations (\ref{dg2}).

 The scale invariant kinetic action of the Higgs field modulus
 can be written in the similar form (\ref{dg1})
\be\label{dg3}
  S_{DG}[g^h]=\int d^4x\frac{\sqrt{-\hat g^h}}{6}~R(\hat g^h), ~~~~~~~\hat
  g^h=\frac{|\Phi|^2}{2}g,
 \ee
  So that the effective inverse Newton coupling constant takes the
  form of the sum of squares of two the scalar fields -- dilaton and
  modulus of the Higgs field
\be\label{dg4}
  ({w}^h)^2=({w})^2-\frac{|\Phi|^2}{2},
 \ee
 After the transformation
\be\label{dg5}
 w=w^h \cosh Q,~~~ \frac{|\Phi|}{\sqrt{2}}=w^h\sinh Q
 \ee
 we get the action of the dilaton GR and SM
\be\label{dg6}
 S_{\rm tot}=S_{DG}[w^h] +
 S_{\rm SM}\left[\frac{|\Phi|}{\sqrt{2}}=w^h\sinh Q\right]
 +\int d^4x (w^h)^2\partial_\mu
 Q\partial_\nu Qg^{\mu\nu}.
 \ee
 The Higgs potential
$
 S_{\rm Higgs}
 \left[\frac{|\Phi|}{\sqrt{2}}=w^h\sinh Q\right]
$
 becomes an superfluous ornamentation, if the field Q begins
 from the  initial data $Q_{I}v$.
The spontaneous symmetry breaking by the initial data is  possible
 due to the Gell-Mann-Oakes-Renner type mechanism
 of the vacuum ordering \cite{252}.

\section{Discussion}

The WMAP observations of the CMB anisotropy now is treated as
distinguishing of one of relativistic  inertial reference frames.
The treatment of the velocity 390 km/c to the Leo as the parameters
of the Lorentz transformations give some requirements to the
fundamentals of the General Theory of Relativity.

The CMB frame is separated from the general coordinate
 transformations by the use of the Fock simplex components,
 where spatial determinant is separated.

The dependence of the energy -- momentum tensor components
  on the determinant component is determined
  by the Lichnerowicz transformations of any field with the conformal
  weight $(n)$  to the conformal-invariant
  (dilaton)  version of GR (\ref{dg1}), where the
  differential element of spatial volume  coincides with
  the coordinate one and the absolute Newton coupling constant is
  converted into the present-day value of dilaton.

The CMB frame is
 invariant with respect to the kinemetric subgroup of
 the general coordinate
transformations, that includes only the reparametrizations of the
coordinate evolution parameter. These reparametrizations means the
existence of the homogeneous time-like dynamics of a whole system of
fields
 that can be treated as the global motion of the universe
 in the {\it field space of events} subgroup of diffeomorphisms of the CMB frame
 can be considered the  foundation of the
 canonical version of the cosmological
 perturbation theory that keeps the the Hamiltonian dynamics with the energy constraint.

 The energy constraint and the Hamiltonian reduction lead to the  definite
 {\it  canonical   rules} of the Universe evolution in the field space of
 events including positive energy of events, the vacuum postulate,
 arrow of geometric time, potential perturbation, which are omitted
 by the standard Lifshits theory.

 It is interesting to apply this canonical approach  to describe the
 fluctuations of CMB temperature.

\begin{flushleft}
\textbf{Acknowledgements}
\end{flushleft}

The authors would like to thanks to A.V. Efremov,
 E.A. Kuraev, V.B. Priezzhev,  and Yu.P. Rybakov
 for interesting and critical
discussions. AFZ is grateful to the National Natural Science
Foundation of China (NNSFC) (Grant \# 10233050)  for partial
financial support. LAG would like to thank to the Bogoliubov--Infeld
program of grants for partial financial support and to Marcin
Cerkaski for very \mbox{interesting} \mbox{discussion}.

\end{document}